\newcommand{\rem}[1]{}

\newcommand{\Id}[1] {\int \! \! {\rm d}^3 #1}

\newcommand{\vr} {{\bf r}}

\documentclass[epj,final]{svjour} 
\usepackage{hyperref}
\usepackage{epsfig}
\begin{document}

\title{
Hund's rule Magnetism in C$_{60}$ ions?
}

\author{Martin L{\"u}ders\inst{1,2,3} \and
        Nicola Manini\inst{1,4,5} \and
        Paolo Gattari\inst{4} \and
        Erio Tosatti \inst{1,2,6}
	}
\institute{ International School for Advanced Studies (SISSA),
Via Beirut 4, 34014 Trieste, Italy
\and
INFM Democritos National Simulation Center, 
and INFM, Unit\`a Trieste, Italy
\and
Daresbury Laboratory, Warrington WA4 4AD, UK 
\and
Dip.\ Fisica, Universit\`a di Milano, Via Celoria 16, 20133
Milano, Italy
\and
INFM, Unit\`a di Milano, Milano, Italy
\and
International Centre for Theoretical Physics (ICTP), 
P.O. Box 586, 34014 Trieste, Italy
}

\date{}

\abstract{
We investigate the occurrence of Hund's rule magnetism in C$_{60}^{n\pm}$
molecular ions, by computing the ground-state spin for all charge states $n$
from $-3$ to $+5$.
The two competing interactions, electron-vibration (e-v, including Jahn
Teller, favoring low spin) and electron-electron (e-e, including Hund-rule
exchange, favoring high spin), are accounted for based on previously
computed ab-initio coupling parameters.
Treating the ion coordinates as classical, we first calculate and classify
the static Jahn-Teller distorted states for all $n$, inclusive of both e-v
and e-e effects.
We then correct the adiabatic result by including the zero-point energy
lowering associated with softening of vibrations at the adiabatic
Jahn-Teller minima.
Our overall result is that while, like in previous investigations, low-spin
states prevail in negative ions, Hund's rule high spin dominates all
positive C$_{60}^{n+}$ ions.
This suggests also that Hund-rule magnetism could arise in fullerene
cation-based solid state compounds, particularly those involving
C$_{60}^{2+}$.
}


\maketitle

\section{Introduction}
Magnetism without transition metals is a potentially exciting subject.
As an example, carbon magnetism has generated some recent interest, in
connection with some ful\-ler\-ene-derived carbon-only magnetic materials
\cite{Makarova01,Wood02}.
In \linebreak[4] 
these materials however the fullerene cage is disrupted into some sort
of three-dimensional bonding network.
In this paper we focus on carbon magnetism that may occur in isolated
fullerene ions, and in ionic fullerene compounds where electrons are added
or subtracted to C$_{60}$ molecules that preserve their overall molecular
integrity.

Unconventional properties, including magnetism, of negative C$_{60}$ 
ions have in fact been previously discussed in the literature.
Due to  the high symmetry of the molecule, 
the threefold degenerate $t_{1u}$ molecular orbital of 
C$_{60}$ will, when partly filled, be affected by Coulomb 
exchange \cite{Martin93,Han00,Lueders02}, which favors 
molecular Hund's rule magnetism, quite similar to that leading to the 
atomic magnetic moments in ordinary $d$ and $f$ elements 
and compounds.
In addition to displaying Hund-rule physics, a high-symmetry molecule 
will however also undergo Jahn-Teller (JT) distortions. The JT
ground state favors electron spin pairing, generally leading 
to larger energy gains for low-spin states than for high-spin states.
Molecular Hund-rule exchange and JT are therefore competing effects
from the point of view of magnetism.
Past studies indicated that in C$_{60}^{n-}$ ions the JT interaction is in
fact somewhat stronger than Coulomb exchange, leading to low-spin ground
states \cite{Lueders02,Bergomi,AMT,MTA,Brouet01}.
%
\footnote{Low-spin
ground state is also observed in a number of diradicals
\cite{Rajca93,Rajca00}, but the situation there is 
quite different from C$_{60}$ ions.
In ideal icosahedral C$_{60}$ ions the degenerate orbitals require a
regular first Hund rule with maximum spin. Density-functional theory (DFT)
calculations do of course confirm that.
A low-spin ground state may be eventually obtained if, upon lowering the
icosahedral symmetry by means of a JT distortion, the corresponding energy
gain happened to be large enough to reverse the undistorted high-spin
situation.
In the aromatic diradicals instead, two distant unpaired spins interact
somewhat weakly through the molecular backbone, and a Heitler-London
singlet ground state appears to be achieved, no distortions involved.
}
Nonetheless, the balance between JT and Hund's rule
is to some degree compound dependent, leading to a close 
match in certain chemical environments. Spin gaps of the order
of 100 meV between the $S=0$ ground state and the $S=1$ excitation 
of the C$_{60}^{2-}$ or equivalently of the C$_{60}^{4-}$ion
were reported in NMR studies of compounds
like K$_4$C$_{60}$ and Na$_2$C$_{60}$, but a narrower 10~meV gap has 
been suggested for the $n=-2$ fluctuating charge state 
in CsC$_{60}$ \cite{Brouet99}.
The magnetism of TDAE-C$_{60}$ \cite{Allemand,Narymbetov} can 
also be attributed to a spin triplet in charge fluctuating
C$_{60}^{2-}$ states \cite{Arovas95}.

The main question which we wish to address here is what ground-state spin
is to be expected for positive C$_{60}^{n+}$ ions. The calculations will be
done in parallel for C$_{60}^{n+}$ and C$_{60}^{n-}$ so as to emphasize the
analogies and the differences that will emerge.  Zero-point effects are
also approximately included, and their effect is shown to be non
negligible.

Data on C$_{60}^{n+}$ molecular ions do not appear to be readily available.
In particular the published photoemission spectra of C$_{60}$
\cite{Bruhwiler97,Canton02} imply C$_{60}^{+}$ final states. This $n=1$
state is affected by JT but clearly not by Hund's rule.
Some charge-transfer compounds contain nominal C$_{60}^{2+}$ ions.  While
it is presently unclear whether higher charge states are accessible,
acceptor compounds with nominal (AsF$_6$)$_2$\linebreak[0]C$_{60}$ and
(SbF$_6$)$_2$C$_{60}$ stoichiometries have been described, containing again
C$_{60}^{2+}$ molecular ions.
Recent data, including anomalously short spin-lattice relaxation 
times \cite{Panich02}, strongly suggest the possibility of magnetism of
the C$_{60}^{2+}$ ions.

Here we address the question of magnetism of C$_{60}^{n+}$ ions by
addressing quantitatively the competition between JT and Coulomb exchange
in these ions.
Both JT and intra-molecular exchange terms are calculated to be 
individually stronger in C$_{60}^{n+}$ than in C$_{60}^{n-}$ 
\cite{Lueders02,Manini01}.
Preliminary calculations and estimates based on the adiabatic 
approximation \cite{Lueders02} did foreshadow the predominance of 
Hund's rule exchange in C$_{60}^{n+}$, as opposed to the predominance
of JT in C$_{60}^{n-}$. 
In the adiabatic approximation however, the JT effect is postulated to be
static, neglecting zero-point motion of the carbon nuclei (responsible for
turning JT from static to dynamic, even at $T=0$).
This simplification might be unsafe, as it is known to underestimate the JT
energy gain \cite{MTA}. In turn, it might seriously impair our
understanding of the competition between high spin and low spin, through
the neglect of very large corrections associated to the different
zero-point vibrational quantum kinetic energy associated with states of
different spin.
Here we estimate these zero-point effects and include them in the coupled
JT-Hund's rule problem, thus correcting the adiabatic result.  An accurate
evaluation of this correction is in fact quite generally a formidable
task. We show here that one lucky feature of C$_{60}^{n+}$ ions is that
their strong e-v coupling makes the basic zero-point correction a 
good approximation to the true result.
Our bottom-line conclusion will be that whereas quantum corrections
to the ground state energy are confirmed to be large, they 
do not reverse the ground-state spin state for any of the C$_{60}^{n+}$ 
ions. In particular the prediction that C$_{60}^{2+}$ should be 
magnetic is maintained. It is in fact reinforced, with the $S=1$ 
ground state about 30~meV lower than the lowest $S=0$ state.

This prediction does not of course imply that compounds containing
C$_{60}^{n+}$ will by necessity be magnetic. Electron kinetic energy
associated with hopping between molecules do in principle favor band
electron spin pairing, and a standard nonmagnetic metallic state. However,
any insulating states that could arise due to very narrow bands, such as a
Mott or more probably a Mott-JT insulator\cite{Fabrizio97}, is very likely
to be magnetic.
In this regard it seems interesting that both (AsF$_6$)$_2$C$_{60}$ and
(SbF$_6$)$_2$C$_{60}$
acceptor intercalated materials, nominally containing doubly positive
fullerene ions
were indeed found to be electrical insulators, with
activation gaps of 0.22 and 0.64 eV respectively \cite{Datars96}.

The starting point of this paper will be the static, adiabatic 
JT-distorted state of all C$_{60}^{n\pm}$ ions treated in the simplest
model Hamiltonians that includes both e-v and e-e interactions.
Each of these static JT wells, or ``valleys'', is characterized by a
reduction of symmetry from icosahedral to some subgroup.
New vibrational frequencies arise at each such valley. We determine these
frequencies by evaluation of the Hessian energy matrix at the 
energy minimum
\cite{ManyModes}.
We use the lowering of the zero-point vibrational energies 
from the undistorted to the JT distorted state to estimate 
the leading quantum correction to the adiabatic approximation.
The next-order quantum correction would arise from weak tunneling between
equivalent valleys. This is the conceptually important step that leads from
static to dynamic JT, with full restoration of the undistorted icosahedral
symmetry.  The associated energy correction expected is however relatively
minor in C$_{60}^{n+}$, in view of the large e-v couplings.
In the limit of infinite coupling the tunneling corrections vanish, and
all quantum effects coincide exactly with the zero-point
lowering.
Therefore, a direct quantitative check that the tunneling corrections
are reasonably small already in the weakest coupling case of C$_{60}^{1-}$,
and already negligible in C$_{60}^{1+}$ justifies us in neglecting
them for all other charge states.
The accuracy of this neglect is, it should be noted, particularly good for
the positive ions C$_{60}^{n+}$, where the e-v coupling is larger than in
C$_{60}^{n-}$.

This paper is organized as follows. Sect.~\ref{model:sec} introduces the
model and the parameters used in this calculation, which is then described
in Sect.~\ref{adiabatic:sec}, along with the properties of the JT valleys for
all values $n$ and $S$.
The zero-point non-adiabatic corrections are described in
Sect.~\ref{nonadiabatic:sec}, and the overall results are 
finally discussed in Sect.~\ref{discussion:sec}.

\section{The model Hamiltonian}

\label{model:sec}

We begin by reviewing here the model Hamiltonian previously introduced in
Ref.~\cite{Lueders02} to describe the electron-vibration coupling and
Coulomb exchange of holes in the $h_u$ fivefold-degenerate highest occupied
molecular orbital (HOMO), and of electrons in the threefold-degenerate
$t_{1u}$ lowest unoccupied molecular orbital (LUMO) of C$_{60}$:
\begin{equation}
\hat{H} = \hat{H}_0 + \hat{T}_{\rm vib}  + \hat{V}_{\rm vib}  + 
\hat{H}_{\rm e-v} + \hat{H}_{\rm  e-e}
\label{modelhamiltonian}
\end{equation}
where 
\begin{eqnarray}
\hat{H}_0     &=& \epsilon \, \sum_{\sigma m}  
\hat{c}^\dagger_{\sigma m} \hat{c}_{\sigma m}, \\
\label{vib-kinetic}
\hat{T}_{\rm vib} &=& \sum_{i\Lambda \mu} \frac{\hbar \omega_{i\Lambda}}{2} 
\hat{P}_{i\Lambda \mu}^2 , \\
\label{vib-potential}
\hat{V}_{\rm vib} &=& \sum_{i\Lambda \mu} \frac{\hbar \omega_{i\Lambda}}{2} 
 \hat{Q}_{i\Lambda \mu}^2  , \\
\label{JT-hamiltonian}
\hat{H}_{\rm e-v} &=& 
\sum_{r\,i\Lambda}
\frac{k^\Lambda g^r_{i\Lambda} \hbar \omega_{i\Lambda}}{2} 
\sum_{\sigma m m' \mu} 
C^{r \Lambda \mu}_{m m'} \, \hat{Q}_{i\Lambda \mu} \,\hat{c}^\dagger_{\sigma m}
\hat{c}_{\sigma\; -m'}  , \\
\hat{H}_{\rm  e-e} &=& 
\frac{1}{2} \sum_{\sigma, \sigma'} \sum_{{m m'}\atop{k k'}} 
w_{\sigma,\sigma'}(m,m';k,k') 
\hat{c}^\dagger_{\sigma m} \hat{c}^\dagger_{\sigma' m'} 
\hat{c}_{\sigma' k'} \hat{c}_{\sigma k}.
\label{Coulomb-hamiltonian}
\end{eqnarray}
are respectively the one-electron Hamiltonian, the vibron kinetic
energy, the harmonic restoring potential toward the equilibrium
configuration of neutral C$_{60}$, the electron-vibron coupling (in the
linear JT approximation) \cite{Manini01,hbyh}, and finally the mutual
Coulomb matrix element representing intra-molecular repulsion 
between the hol\-es/el\-ectrons \cite{Lueders02}.
Here $\hat{c}^\dagger_{\sigma m}$ denote creation operators of either a
hole in the HOMO or an electron in the LUMO, described by the
single-particle wave functions $\varphi_{m\sigma}(\vr)$.
$\sigma$ indicates the spin projection; $m$ labels the component within the
degenerate electronic multiplets, according to the $C_5$ character from the
$I_h\supset D_5\supset C_5$ group chain
\cite{hbyh,Butler81}.
$i$ enumerates the vibration modes of symmetry $\Lambda$ (2 $A_g$, 8 $H_h$, and 6
$G_g$ modes, the latter being JT active in the hole case only).
$C^{r \Lambda \mu}_{m m'}$ are Clebsch-Gordan coefficients \cite{Butler81}
of the icosahedral group $I_h$, for coupling  $h_u$/$t_{1u}$ states to
phonons of symmetry $\Lambda$.
$r$ is a multiplicity label, relevant for $h_u$ holes and $H_g$ vibrations
only, where it takes two values, 1 and 2 \cite{Manini01,Butler81}.
$\hat{Q}_{i\Lambda \mu}$ are the dimensionless molecular normal-mode
vibration coordinates (measured from the adiabatic equilibrium
configuration of neutral C$_{60}$, in units of the length scale
$x_0(\omega_{i\lambda})=\sqrt{\hbar/ (\omega_{i\lambda} \, m_{\rm C})}$
associated with each harmonic oscillator where $m_{\rm C}$ is the mass of the C
atom), and $\hat{P}_{i\Lambda \mu}$ the corresponding conjugate momenta.
Finally, spin-orbit, exceedingly small in C$_{60}$ \cite{TMG}, is 
neglected throughout.

\begin{table}
\caption{\label{e-v-params:table}
Computed vibrational eigenfrequencies, and e-v linear coupling 
parameters for the $h_u$ HOMO and $t_{1u}$ LUMO in C$_{60}$
\cite{Manini01}.
}
\begin{center}
\begin{tabular}{rrccc}
\hline
\hline
$\hbar\omega_{i\Lambda}$  &       $\hbar\omega_{i\Lambda}$  &
$g^1_{i\Lambda}$    &       $g^2_{i H_g}$       
&       $g_{i\Lambda}$\\
cm$^{-1}$               &       meV     &   (HOMO)    &  (HOMO)
&       (LUMO)\\
\hline
$A_g$\quad\quad\quad\\
500     &       62.0    &       0.0591   &       -       &      0.1565 \\
1511    &      187.4    &       0.2741   &       -       &      0.3403 \\
\hline
$G_g$\quad\quad\quad\\
483     &       59.9    &       0.7567   &       -       &      - \\
567     &       70.3    &       0.1024   &       -       &      - \\
772     &       95.7    &       0.8003   &       -       &      - \\
1111    &      137.8    &       0.6239   &       -       &      - \\
1322    &      163.9    &       0.2277   &       -       &      - \\
1519    &      188.4    &       0.4674   &       -       &      - \\
\hline
$H_g$\quad\quad\quad\\
261     &      32.4     &       3.0417   &    $-0.0045$  &      0.4117 \\
429     &      53.2     &       1.0587   &\ \ \,0.6131   &      0.4886 \\
718     &      89.0     &       0.0103   &\ \ \,0.9950   &      0.3500 \\
785     &      97.3     &       0.7836   &    $-0.0309$  &      0.2238 \\
1119    &     138.7     &       0.0514   &\ \ \,0.2151   &      0.1930 \\
1275    &     158.0     &       0.4586   &\ \ \,0.2440   &      0.1382 \\
1456    &     180.5     &       0.8482   &\ \ \,0.4530   &      0.3152 \\
1588    &     196.9     &       0.7436   &    $-0.4488$  &      0.2893 \\
\hline
\hline
\end{tabular}
\end{center}
\end{table}

\begin{table}
\caption{\label{e-e-params:table}
The Coulomb parameters for C$_{60}^{n\pm}$, defining $\hat{H}_{\rm e-e}$
through (\ref{Coulomb-hamiltonian}) and (\ref{Coulomb-ints}), as obtained
from the DFT calculations of Ref.~\cite{Lueders02}.
%
}
\begin{center}
\begin{tabular}{lcr}
\hline
\hline
&Parameter & Value\\
&        &       [meV]  \\
\hline
HOMO (C$_{60}^{n+}$) \\
&$F_1$   &15646 \\
&$F_2$   &  105 \\
&$F_3$   &  155 \\
&$F_4$   &   47 \\
&$F_5$   &    0 \\
&$U$     & 3097 \\
\hline
LUMO (C$_{60}^{n-}$) \\
&$J$     &   32 \\
&$U$     & 3069 \\
\hline
\end{tabular}
\end{center}
\end{table}

For holes, the JT model defined by Eq.~(\ref{vib-kinetic}), (\ref{vib-potential}),
(\ref{JT-hamiltonian}), for C$_{60}^{n+}$ is conventionally denoted as
$h^n \otimes (A + G + H)$: $h^n$ refers to the hole occupancy of the $h_u$
HOMO, and $A$, $G$, $H$ refer to the 2 nondegenerate $A_g$, 6
fourfold-degenerate $G_g$ and 8 fivefold-degenerate $H_g$ molecular
vibration modes that are linearly coupled to $h_u$ in icosahedral
symmetry \cite{Manini01,hbyh,CeulemansII}.
For electrons, the JT model is $t^n \otimes (A + H)$, $t^n$ 
referring to $n$ electrons occupying the $t_{1u}$ LUMO, linearly 
coupled to the 2 $A_g$ and 8 $H_g$ vibrational modes only.
In all calculations we shall adopt the numerical values of the e-v coupling
parameters $g^r_{i\Lambda}$, listed in Table~\ref{e-v-params:table},
previously obtained from first-principles Density DFT electronic structure
calculations in Ref.~\cite{Manini01}.
A recent DFT calculation \cite{Saito02} based on a different functional
reported couplings that are similar on the whole to those of
Table~\ref{e-v-params:table}, the main difference concerning 
a closer competition between $D_{3h}$ and $D_{5d}$ valleys for
the distortion of C$_{60}^{1-}$.
The JT stabilization energy based on the DFT parameters of
Ref.~\cite{Manini01} is only about one fifth of that found based on the
intermediate neglect of differential overlap (INDO) model \cite{Bendale92}.
%
We think that these earlier calculations, as well as more recent
Hartree-Fock (HF) estimates \cite{gosia03} also suggesting large e-v
couplings and energy gains are somewhat less dependable. We believe the
smaller DFT JT gains more realistic for two main reasons: (i) the HF
calculations miss an important loss of correlation energy due to opening of
the JT gap; (ii) preliminary results indicate that the HOMO photoemission
spectrum based on the DFT parameters~\cite{MGTunpublish} is in very
good agreement with experiment \cite{Bruhwiler97,Canton02}.
By contrast, for C$_{60}^-$, DFT computed electron-phonon
couplings~\cite{vzr,Schl,Antropov,Green96} 
appear to be significantly smaller than those obtained from photoemission
of C$_{60}^-$~\cite{Gunnarsson}.
While we can offer no explanation for that discrepancy in negative ions, we
note that the DFT-derived couplings used here for C$_{60}^{n+}$ are
probably also slightly underestimated, however probably only by some 10 to
20\%.

The e-v couplings $g^r_{i\Lambda}$ in Eq.~(\ref{JT-hamiltonian}) are
dimensionless, measured in the units of the corresponding harmonic
vibrational energy quantum $\hbar\omega_{i\Lambda}$.
Modes (e.g.\ the second $G_g$ mode) with $g^r_{i\Lambda}\ll 1$ 
are thus only weakly coupled; conversely,
modes (e.g.\ the lowest $H_g$ mode) with a large $g^r_{i\Lambda} > 1$ 
are strongly e-v coupled.
The numerical factors $k^{A_g}=5^{\frac 12}$,
$k^{G_g}=\left(5/4\right)^{\frac 12}$, $k^{H_g}=1$ for the HOMO, and
$k^{A_g}=3^{\frac 12}$, $k^{H_g}=6^{\frac 12}$ for the LUMO are introduced
for compatibility with
the normalization of the e-v parameters in Ref.~\cite{Manini01}.

The Coulomb matrix elements are defined by:
\begin{eqnarray}
\label{Coulomb-ints}
w_{\sigma,\sigma'}(m,m';k,k') = \Id{r} \! \Id{r'}& \, \\\nonumber
\varphi^{*}_{m \sigma}(\vr) \, 
\varphi^{*}_{m'\sigma'}(\vr') \,
u_{\sigma,\sigma'}(\vr,\vr') &
\varphi_{k\sigma}(\vr) \,
\varphi_{k'\sigma'}(\vr') 
\end{eqnarray}
where $u_{\sigma,\sigma'}(\vr,\vr')$ is an effective Coulomb repulsion,
screened by all other electrons of the molecule (eventually also
by electrons in all other molecules in a solid state compound; but
we shall focus here on the isolated ion).
A detailed symmetry analysis \cite{Lueders02} shows that, assuming
spin-independence of the orbitals, this set of
coefficients can be expressed as 
\begin{equation}
w_{\sigma,\sigma'}(m,m';k,k') = \sum_{r,r',\Lambda} F^{r,r',\Lambda}
\left( \sum_\mu C^{r \Lambda \mu}_{m k} \, 
C^{r' \Lambda \mu}_{m' k'} \right) 
\label{Fdecomposition}
\end{equation}
in terms of a minimal set of independent Slater-type parameters
$F^{r,r',\Lambda}$ \cite{Cowan}.
A DFT estimate of these parameters was previously obtained in
Ref.~\cite{Lueders02}, and for our calculation we adopt those values of
the Coulomb parameters. They are reproduced for completeness in
Table~\ref{e-e-params:table}.
%
%
Other sets of couplings, obtained by means of a simple (and clever)
model for the HOMO and LUMO orbitals \cite{Nikolaev02} and by a fit to
multi-configuration HF calculations \cite{gosia03},
are in substantial agreement with each other, but they are both much larger
than the DFT couplings of Ref.~\cite{Lueders02}, due to complete neglect of
screening.
We believe that the actual Coulomb parameters of C$_{60}$ lie somewhere in
between the DFT couplings used here and the ``bare'' ones of
Ref.~\cite{Nikolaev02,gosia03}, but most likely closer to the DFT ones, due
to the large polarizability of C$_{60}$.

For the HOMO Coulomb parameters we use the shorthands
\begin{eqnarray}
F_1= F^{A_g},\
F_2= F^{G_g},\
F_3= F^{1,1,H_g}, \\ \nonumber
F_4= F^{2,2,H_g},\
F_5= F^{1,2,H_g}.
\label{fnumbering}
\end{eqnarray}
The Coulomb energy connected with the total molecular charge
fluctuation in the HOMO, conventionally called the hole 
``Hubbard U'' is given by the combination
\begin{equation}
U = \left( \frac{F_1}{5} - \frac{4 \, F_2}{45} 
- \frac{F_3}{9} - \frac{F_4}{9} \right) .
\label{Udefinition}
\end{equation}
For electrons in the LUMO, we have instead an electron
Hubbard U given by $U=F^{A_g}/3 - F^{H_g}/3$ and a Hund-rule
exchange $J= F^{H_g}/2$.

In either case, of electron or of holes, $U$ defines an 
average Coulomb repulsion within the multiplet of states of that
molecular ion C$_{60}^{n\pm}$, such that the average energy for each $n$
\begin{equation}
E^{\rm ave}(n)=
\epsilon  \, n + U \frac {n (n-1)}2 \, .
\label{Eave:eq}
\end{equation}
We observe that $U$ differs from a  more common definition of the Hubbard
repulsion in lattice models, which involves the lowest state in each
$n$-configuration: $ U^{\rm min} = E^{\rm min}(n+1) + E^{\rm min}(n-1) - 2
E^{\rm min}(n)$.  This second definition is inconvenient here, since it
depends wildly on $n$.
The Hubbard U's are given here for completeness, and because
they will be useful in different contexts. However we must recall
that the $U$ term is irrelevant for the determination of the
ground-state spin of each C$_{60}^{n\pm}$ ion.

\section{The adiabatic JT valleys}
\label{adiabatic:sec}

Though a rather idealized
description of real C$_{60}^{n\pm}$ ions, the model 
Hamiltonian (\ref{modelhamiltonian}), does nevertheless
not lend itself to an exact solution.
The vibronic eigenstates are generally complicated combinations 
in the direct product
of each (up to 252-dimensional) fixed-$n$ electronic space, times the
infinite-dimensional space of the vibrational degrees of freedom.
Even with the help of spin and orbital symmetries, exact solutions of the
quantum problem (\ref{modelhamiltonian}) are only available in two limiting
cases, namely the limit of weak e-v coupling \cite{MTA,Reno,Delos96} and
that of infinitely strong e-v coupling
\cite{AMT,ManyModes,CeulemansII,Moate97}.
For C$_{60}^{n\pm}$, where the couplings range from 
intermediate to large, some
approximations are therefore called for.
If the phonon kinetic term $\hat{T}_{\rm vib}$ is neglected in
the so-called adiabatic approximation, the
distortion operators $\hat{Q}_{i\Lambda \mu}$ are replaced by 
c-number coordinates.
This approximation yields the leading ground-state energy lowering $\propto
\sum (g^r_{i\Lambda})^2 \hbar\omega_{i\Lambda}$, exact 
in the limit of large e-v couplings $g^r_{i\Lambda}\to \infty$. 
[Note however that, in this limit, both initial 
assumptions of harmonic vibrations (\ref{vib-potential})
and of linear e-v coupling (\ref{JT-hamiltonian}) become
anyway questionable].
In Sect.~\ref{nonadiabatic:sec} we shall deal with the leading quantum
corrections to the adiabatic approximation $\propto\sum (g^r_{i\Lambda})^0
\hbar\omega_{i\Lambda}$ by taking zero-point energy shifts into account.

The classical treatment of the vibration coordinates breaks 
the full molecular symmetry (here icosahedral symmetry) in all 
configurations possessing a non\-zero distortion $Q_{i\Lambda \mu}$
($\Lambda = G_g,H_g$). Therefore states of different icosahedral 
symmetry representations are in general intermixed, leaving only 
the total number of holes $n$, the total spin $S$ and its 
projection $S_z$ conserved in the adiabatic ground state.
Here we will assume the orbitals to remain basically unchanged upon JT
distortion, thus neglecting any small JT-induced change of the Coulomb
Hamiltonian $\hat{H}_{\rm e-e}$.  The latter is therefore still determined
according to
Eqs.~(\ref{Coulomb-hamiltonian},\ref{Coulomb-ints},\ref{Fdecomposition}) by
the same parameters $F_i$ of Table~\ref{e-e-params:table}, as in the
undistorted icosahedral configuration.
We will also assume no change of the vibration frequencies 
$\omega_{i\Lambda}$ and couplings $g^r_{i\Lambda}$ upon charging.
While this is of course at variance with much established evidence
showing vibration frequency shifts in the percent range
per each added electron, it is perfectly in line with the
other approximations intrinsic in our model, and in its 
solution. 
We leave the $A_g$ modes out of the adiabatic calculation, 
since despite their nonzero linear e-v coupling they simply
contribute a trivial spin- and symmetry-independent 
term
\begin{eqnarray}
E^{A_g}(n) = -\frac 18 n^2 \sum_i g_{i A_g}^2 \hbar \omega_{i A_g}
&=& -a\, n^2;  \\ \nonumber
 a=1.79~{\rm meV}\ {\rm (HOMO)}, \quad 
 a&=& 2.90~{\rm meV}\ {\rm (LUMO)},
\label{ag_energy}
\end{eqnarray}
to the total energy $E^{\rm ave}(n)$ [Eq.~(\ref{Eave:eq})], which could effectively be
included into $U$.
Because of particle-hole symmetry (exchanging creation and annihilation
operators) of the Hamiltonian (excluding $H_0$, the $A_g$ modes and the
average $U$ contribution), positive charges $n>5$ can always be mapped onto
$n\leq 5$, and negative $n>3$ onto $n\leq 3$.

\begin{table*}
\caption{\label{energies:tab}
The three contributions (restoring harmonic potential, e-v coupling, and
e-e exchange -- the $\left[U n(n-1)/2\right]$ term is excluded from
$\hat{H}_{\rm e-e}$) in the adiabatic energy $E_{\rm adiab}$ for
C$_{60}^{n\pm}$ in all possible charge and spin states.
As usually occurs in JT, the contribution $\left\langle \hat{H}_{\rm e-v}
\right\rangle = -2 \left\langle \hat{V}_{\rm vib} \right\rangle =-2 V_{\rm
vib}({\bf Q}_{\rm min})$.
The following columns report the leading zero-point energy correction
$E_{\rm zero}$ [Eq.\ (\ref{Ezero})] and the sum of this correction
to the adiabatic energy, representing our best estimate of the 
energy of each valley, relative to the undistorted state of the ion.
For C$_{60}^{\pm}$, the last column reports the exact energy obtained by
Lanczos diagonalization.
All energies are in meV. Boldface indicates the ground-state total energy
at each $n$ (high spin for cations, low spin for anions).
For positive ions, comparison with the exact result in the last column
indicates a fairly good approximation already for $n$ =1; and we expect the
accuracy to be even better for $n$ = 2,3,4,5 holes.
}
\begin{center}
\begin{tabular}{cc|rrrrrcc}
\hline
\hline
$n$&$S$  &$\left\langle \hat{V}_{\rm vib} \right\rangle$ 
                & $\left\langle \hat{H}_{\rm e-v}\right\rangle$
                        & $\left\langle\hat{H}_{\rm e-e}\right\rangle$
                                &$E_{\rm adiab}$
                                        & $E_{\rm zero}$
                                                &$E_{\rm adiab}+E_{\rm
                zero}$
                                                  &$E_{\rm exact}$\\
\hline
\multicolumn{2}{c|}{C$_{60}^{n+}$} & \\
1       &       1/2     &       69      &       $-$138  &       0       &       {\bf $-$69}   &       $-$51   &       {\bf $-$120} & $-$103 \\
 & \\
2       &       0       &       270     &       $-$540  &       141     &       $-$129  &       $-$72   &       $-$201  \\
        &       1       &       99      &       $-$197  &       $-$43   &       {\bf $-$142}  &       $-$92   &       {\bf $-$234}  \\
 & \\
3       &       1/2     &       267     &       $-$534  &       99      &       $-$168  &       $-$122  &       $-$290  \\
        &       3/2     &       99      &       $-$197  &       $-$123  &       {\bf $-$222}  &       $-$92   &       {\bf $-$314}  \\
 & \\
4       &       0       &       361     &       $-$723  &       162     &       $-$200  &       $-$133  &       $-$332  \\
        &       1       &       229     &       $-$459  &       19      &       $-$211  &       $-$134  &       $-$345  \\
        &       2       &       69      &       $-$138  &       $-$238  &       {\bf $-$308}  &       $-$51   &       {\bf $-$359}  \\
 & \\
5       &       1/2     &       308     &       $-$616  &       105     &       $-$203  &       $-$150  &       $-$353  \\
        &       3/2     &       169     &       $-$338  &       $-$87   &       $-$256  &       $-$98   &       $-$354  \\
        &       5/2     &       0       &       0       &       $-$397  &       {\bf $-$397}  &       0       &       {\bf $-$397}  \\
\hline
\multicolumn{2}{c|}{C$_{60}^{n-}$} & \\
1       &       1/2     &       38      &       $-$77   &       0       &
 {\bf $-$38}   &       $-$113  &       {\bf $-$152}  &  $-$76\\
 & \\
2       &       0       &       149     &       $-$298  &       56      &       {\bf $-$93}   &       $-$115  &       {\bf $-$207}  \\
        &       1       &       38      &       $-$77   &       $-$32   &       $-$71   &       $-$113  &       $-$184  \\
 & \\
3       &       1/2     &       113     &       $-$225  &       28      &       $-$85   &       $-$171  &       {\bf $-$256}  \\
        &       3/2     &       0       &       0       &       $-$97   &       {\bf $-$97}   &       0       &       \,\ $-$97       \\
\hline
\hline
\end{tabular}
\end{center}
\end{table*}

For each charge $n$ and spin $S$ and as a function of the vibration coordinates
${\bf Q}$, the lowest adiabatic potential surface
$V^{\rm ad}_{n,S}({\bf Q})$ is obtained as the sum of the lowest eigenvalue of
$\hat{H}_{\rm e-e} + \hat{H}_{\rm e-v}({\bf Q}$) in the $n$-electron
spin-$S$ sector, plus the harmonic restoring term $V_{\rm vib}({\bf Q})$.
%
Allowing the 64
($6\times 4$ $G_g$ \footnote{The $G_g$ modes are irrelevant for the 
negative ions, but relevant for positive ions} plus $8\times 5$ $H_g$.) 
vibration coordinates ${\bf Q}$
to relax, we determine the optimal distortions ${\bf Q}_{\rm min}$ by full
minimization of $V^{\rm ad}_{n,S}({\bf Q})$ in the space of all vibration
coordinates ${\bf Q}$.
Details of this minimization were previously reported in
Ref.~\cite{Lueders02,Leuven02}.

Table~\ref{energies:tab} summarizes the optimal adiabatic energies
\linebreak[4]
$E_{\rm adiab}(n,S)=V^{\rm ad}_{n,S}({\bf Q}_{\rm min})$. The separate
vibration, e-v, and exchange contributions are specified.
The results show that the adiabatic valley minima of positive 
C$_{60}$ ions are systematically lower for high-spin
states, while the adiabatic outcome for negative ions 
favors marginally low spin, and is essentially uncertain,
with small and poorly reliable energy differences between 
high- and low-spin states.

Positive and negative ions differ qualitatively also 
in the geometric nature of the JT distortions characterizing the minimal
valleys in the space of coordinates ${\bf Q}$. 
The $t^n\otimes H$ model applicable to electrons in C$_{60}^{n-}$ leads 
to a flat continuous manifold of equivalent points. The 
valleys are really troughs, which could be imagined as  ``Mexican hats''
\cite{AMT,ob72,ob96};
the study of these 2-dimensional ($n=1,2$) and 3-dimensional ($n=3$)
troughs is a classic topic in JT physics \cite{ob96,ChoB97}.

Conversely, for holes in C$_{60}^{n+}$ the $h^n\otimes (G + H)$ JT model
gives rise to discrete sets of isolated minima
\cite{Manini01,hbyh,CeulemansII,Moate96}.
The nature and symmetry of these discrete valleys depends on
the details of the e-v couplings and on the interplay with Coulomb
interaction.
We shall describe here in some detail the minima obtained 
in this calculation.

To identify these minima, we generate about a hundred randomly distributed
distortions away from the $I_h$ high-symmetry point, and let the vibration
coordinates relax from there to the closest energy minimum, by combined
standard (simplex and conjugate-gradients) minimization algorithms.
We then apply the symmetry operations of the icosahedral group to the each
minimum found, locating all possible equivalent minima.
Although the method employed is not exhaustive, the application of symmetry
and the thoroughness of our search, carried out with a large variety of
starting points guarantees in practice the retrieval of all relevant
minima.
We can readily discard, for example, the few cases where minimization led
to saddle points or to secondary, local minima, based on simple comparison
of the energy values.
In this way, for each $n$ and $S$, we finally obtain the set of equivalent
valleys that are global adiabatic energy minima for C$_{60}^{n+}$.
The amount of ``radial'' distortion at the adiabatic minima for each mode
is tabulated in the Appendix.

\begin{table*}
\caption{
\label{tab-symm}
Number and local symmetry of the JT minimal valleys for each charge $n$ and
spin $S$ in C$_{60}^{n\pm}$ ions.
The 5th column gives, for each given valley, the number of equivalent
valleys that are first, second, etc.\ neighbors of that valley in ${\bf Q}$
space.
The last column gives the total magnitude of the dimensionless 
JT distortion at each minimum.
}
\begin{center}
\begin{tabular}{cc|ccll}
\hline
\hline
$n$     &$S$    &       number of &     local   &       number of       &       distortion      \\
        &       &       minima    &     symmetry&       1$^{\rm st}$, 2$^{\rm nd}$, 3$^{\rm rd}$, 4$^{\rm th}$... neighbor minima   &       $\left|{\bf Q}_{\rm min}\right|$        \\
\hline
\multicolumn{2}{c|}{C$_{60}^{n+}$} & \\
1       &       1/2     &       6       &$D_{5d}$       &       5              &       1.58\\
& \\
2       &       0       &       6       &$D_{5d}$       &       5                       &       3.12\\
        &       1       &       15      &$D_{2h}$       &       4  4  4  2      &       1.87\\
& \\
3       &       1/2     &       30      &$C_{2h}$       &       2 1 2 4 4 2 2 2 4 4 2   &       3.08\\
        &       3/2     &       15      &$D_{2h}$       &       4  4  4  2              &       1.87\\
&\\
4       &       0       &       10      &$D_{3d}$       &       3  6            &       3.52\\
        &       1       &       30      &$C_{2h}$       &       2 2 2 2 1 4 4 4 6 2     &       2.85\\
        &       2       &       6       &$D_{5d}$       &       5                       &       1.58\\
& \\
5       &       1/2     &       60      &$C_{2h}$       &       1 2 2 4 4 4 2 2 2 4 2 \, 2 4 2 2 2 4 4 4 2 2 1 1  &       3.27\\
        &       3/2     &       30      &$C_{2h}$       &       8  12 8  1              &       2.46\\
        &       5/2     &       1       &$I_h$          &       0               &       0\\
\hline
\multicolumn{2}{c|}{C$_{60}^{n-}$} & \\ 
1       &       1/2     &       $\infty$& $C_i$ or higher &     $\infty$        &       0.91\\
& \\
2       &       0       &       $\infty$& $C_i$ or higher &     $\infty$        &       1.79\\
        &       1       &       $\infty$& $C_i$ or higher &     $\infty$        &       0.91\\
& \\
3       &       1/2     &       $\infty$& $C_i$ or higher &     $\infty$        &       1.55\\
        &       3/2     &       1       & $I_h$           &     0               &       0\\
\hline
\hline
\end{tabular}
\end{center}
\end{table*}

In Table~\ref{tab-symm} we summarize some global properties of the 
JT energy valleys  for C$_{60}^{n\pm}$ in all spin sectors.
In these many-mode JT systems, the local symmetry of an optimal distortion
is described in terms of the subgroup $G_{\rm local}\subset I_h$ of
symmetry operations which leave that minimum invariant.
The minima in the simple $n=1$ $S=\frac 12$ case, where e-e 
interactions are irrelevant, are believed to be six valleys 
of $D_{5d}$ symmetry \cite{Manini01,hbyh,CeulemansII} -- the 
alternative possibility of ten $D_{3d}$ valleys apparently 
disfavored by the specific couplings obtained for fullerene \cite{Manini01}.
Table~\ref{tab-symm} reveals a symmetry
between the minima for $n=4$ $S=2$ and $n=1$ $S=\frac 12$, as well as
for $n=2$ $S=1$ and $n=3$ $S=\frac 32$. These symmetries are at 
first sight surprising, but can be readily explained by applying a
particle-hole transformation to the fermion operators of only one spin
kind, in the fully spin-polarized states. This transformation maps the
Hamiltonian matrices of an $n$-particle states (apart from a constant
exchange term) into those of $(5-n)$ particle states, with a sign change of
the vibron interaction $\hat{H}_{\rm e-v}$ that shifts each minimum ${\bf
Q}_{\rm min}$to the opposite locations $-{\bf Q}_{\rm min}$.
We have therefore $V^{\rm ad}_{4,2}({\bf Q}) = V^{\rm ad}_{1,1/2}(-{\bf Q})
+ C$, with $C = -\frac 23 F_2 - \frac 56 F_3 -\frac 56 F_4=-238$~meV.
The same connection between the $n=2$ $S=1$ and the $n=3$
$S=\frac 32$ adiabatic potentials involves an energy shift $C= -\frac{2}{9}
F_2 - \frac{5}{18} F_3 - \frac{5}{18} F_4 =-80$~meV.
The same symmetry relates $n=5$ $S=\frac 52$ to $n=0$ $S=0$, where no JT
distortion takes place.

In general, the number of JT valleys (third column of Table~\ref{tab-symm})
is related to the local symmetry. It is basically given by the ratio
$|I_h|/|G_{\rm local}|$ of the order of the icosahedral group (120) to
the order of the invariant subgroup $G_{\rm local}$.
However, $n=5$ holes make an exception to this rule.
Here, at half filling, in addition to the $I_h$ symmetry, the system is
particle-hole symmetric, i.e.\ invariant under exchange of creation and
annihilation operators of both spin kinds.
This transformation leaves the Coulomb Hamiltonian $\hat{H}_{\rm e-e}$
invariant, and again changes sign of the vibronic interaction $\hat{H}_{\rm
e-v}$:
hence, given any minimum ${\bf Q}_{\rm min}$, its opposite $-{\bf
Q}_{\rm min}$ is also an equivalent minimum of the potential 
energy surface.
For $S=\frac 12$, this leads to a doubling of the minima: the local
$C_{2h}$ symmetry would lead to 30 minima, but 30 additional 
equivalent minima
are added in the opposite positions by particle-hole symmetry.
For $S=\frac 32$ instead, the number of minima remains 30, 
since for each
minimum there is one of the $I_h$ symmetry operations, a $C_2$ rotation,
that transforms this minimum into its opposite 
point\footnote{The real-space inversion, acting on the real-space
displacements, leaves all JT-active normal-mode coordinates
invariant, since they are even ($g$) under inversion.}.

Table~\ref{tab-symm} contains also some information about the connectivity
of the minima in ${\bf Q}$ space.
In many cases, the specification of the number of first, second,
etc.\ neighbors of a given minimum is sufficient to clarify completely the
topology of the minima in the 64-dimensional space.
In particular, the $D_{5d}$ wells of the $n=1$ $S=\frac 12$, of the $n=2$
$S=0$, and of the $n=4$ $S=2$ surfaces are located on the six vertexes of
five-dimensional regular simplexes, generalizations of the 3D tetrahedron,
each minimum being equidistant to all the others.
In analogy, the connectivity of the 10 $D_{3d}$ minima for $n=4$ $S=0$ is
the same as that depicted in Fig.~1b of Ref.~\cite{hbyh} for a different
situation.
For the other cases of lower symmetry, the number of neighbors of any given
order must be complemented by some extra connectivity information, for
which we refer to previous work \cite{Leuven02}.
We only observe that for $n=2$ $S=1$ (equivalently for $n=3$ $S=\frac 32$),
each of the 15 minima is linked to four nearest-neighbor minima, which, in
turn, are linked to more minima, forming a completely connected regular
polytope.
%
For $n=3$ $S=\frac 12$ and $n=4$ $S=1$, the 30 minima are divided into 6
pentagonal ``clusters'' of five nearest-neighboring minima.
In contrast, for $n=5$ $S=\frac 12$, nearest-neighbor wells come in pairs.
Finally, the 30 $C_{2h}$ minima for $n=5$ $S=\frac 32$, show the largest
connectivity, and sit at the vertexes of a highly symmetric polytope.

\begin{figure*}
\centerline{
\epsfig{file=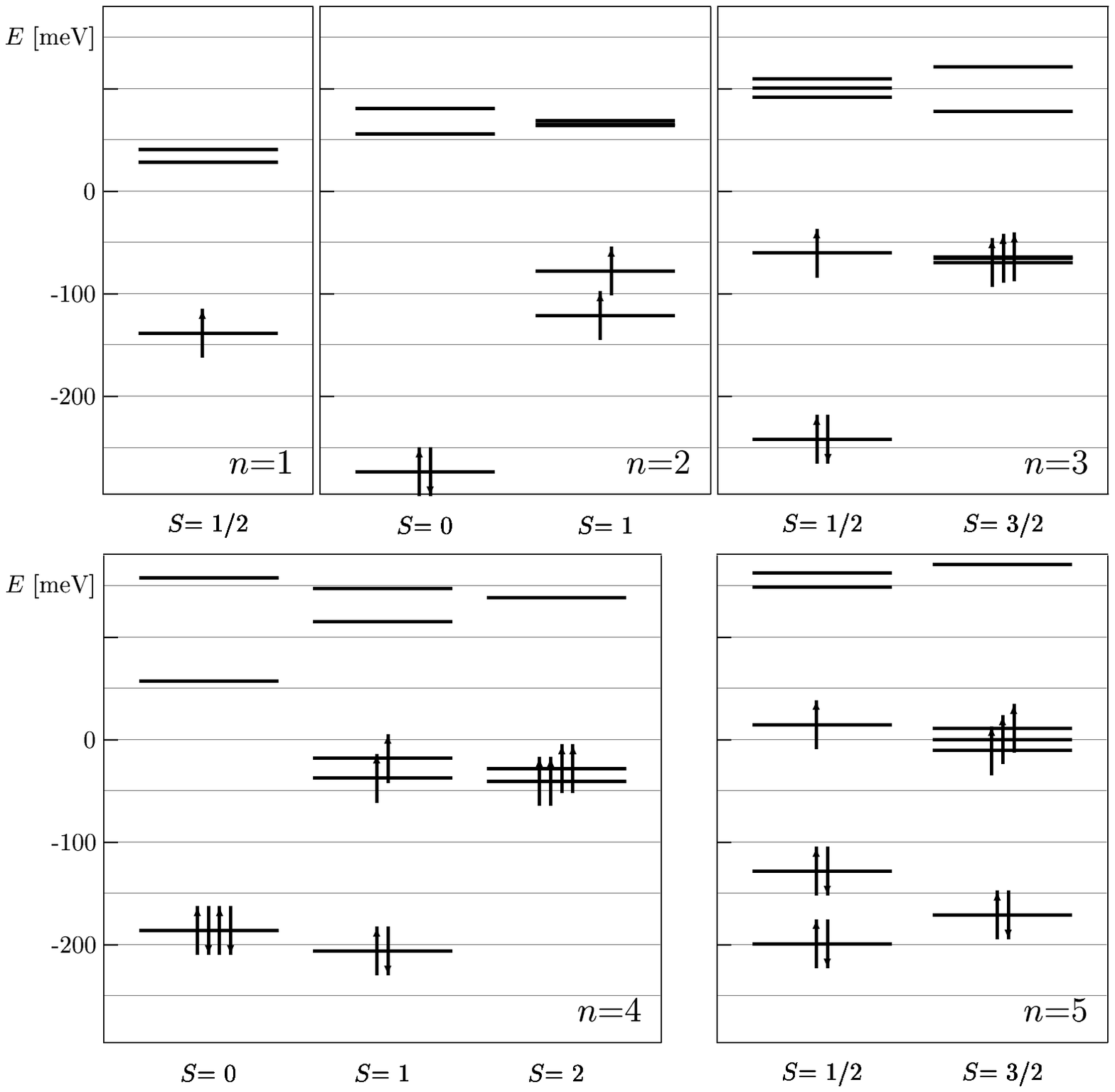,width=12cm}
}
\caption{\label{elfilling:fig}
A line-spectrum representation of the dominant single-determinant
electronic state in the optimal adiabatic configuration of
C$_{60}^{n+}$ for each spin $S$.  Levels are ordered for holes.
}
\end{figure*}

Figure~\ref{elfilling:fig} pictures the electronic state of the holes at
the adiabatic valleys of C$_{60}^{n+}$.
As typical of JT, distortions are such that relatively large gaps open
between empty and filled levels.
For $n>1$ this single-determinant picture is only approximate: it
represents the dominant uncorrelated configuration within the fully
correlated exact electronic ground state at each well.

The adiabatic valleys described above are the joint result of e-v couplings
and of e-e Coulomb interactions. One can understand better the 
specific role of Coulomb interaction for the JT distortion
by comparing the number and symmetry of the valleys described
above to those obtained by the same procedure in a hypothetical
noninteracting case, obtained by setting all $F_j=0$.
Doing that, we find basically the same general picture of valleys for $n=1$
(obviously), but also for $n=2$, $n=3$, $n=4$ $S=1,\,2$ and $n=5$ $S=\frac
32,\,\frac 52$.
In all remaining cases the e-e correlations also introduces a qualitative
change in the topology and local symmetry of the minima.
Without interactions, 15 $D_{2h}$ minima would replace the 10 $D_{3d}$ minima for
$n=4$ $S=0$; and 20 $D_{3d}$ minima replace the 60 $C_{2h}$ minima for
$n=5$ $S=1/2$.
In all cases (except of course $n=1$ and the symmetric $n=4$ $S=2$), the
amount of distortion is a few per cent larger in the uncorrelated case than
when exchange is included, as expected from competing interactions.
For example, without correlation, the total distortion for $n=2$ $S=0$
would be 3.16, i.e.\ exactly twice the distortion for $n=1$, as routinely occurs in
JT when exchange plays no role.

\section{Zero-point quantum corrections}
\label{nonadiabatic:sec}

\begin{figure}
\centerline{
\epsfig{file=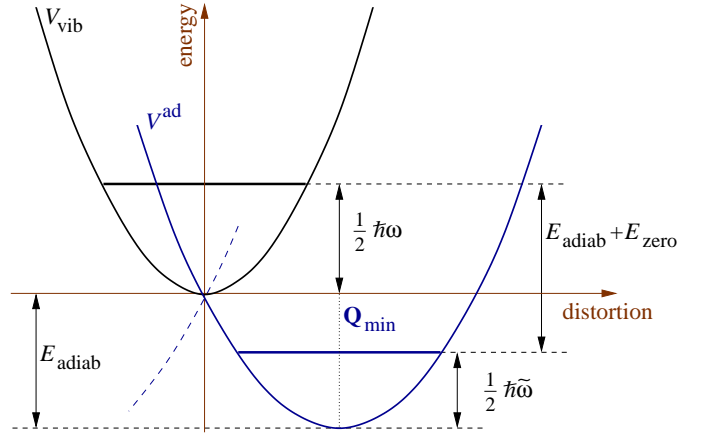,width=8.9cm}
}
\caption{\label{zeroEnergetics:fig}
A pictorial illustrating the origin of the zero-point energy gain due to
vibrational softening at the JT well.
The horizontal axis represents a generic distortion coordinate, the
horizontal thick lines represent the quantum ground states in the non-JT
and JT potential wells.
}
\end{figure}

The vibrational quantum kinetic energy, a quantity neglected at the
adiabatic level, does in fact contribute quite substantially to the JT
energetics of these multi-mode systems characterized by intermediate
couplings.
The zero-point energy gain associated to the softening of the vibrational
frequency at the JT-distorted minima is illustrated in
Fig.~\ref{zeroEnergetics:fig}. It represents the leading quantum correction
to the static JT energetics \cite{ManyModes,Leuven02,LeuvenError}.
To compute this correction for each charge $n$ and spin $S$, by finite
differences we evaluate the Hessian matrix of the second-order derivatives
of the lowest adiabatic potential sheet, at each of the adiabatic JT
minima:
\begin{equation}
{\cal H}(n,S)_{\alpha\, \alpha' }=
\left. \frac {\partial^2 V^{\rm ad}_{n,S}({\bf Q})}
{\partial Q_{\alpha}\, \partial Q_{\alpha'}}
\right|_{{\bf Q}_{\rm min}(n,S)}
\end{equation}
where $\alpha$ is a collective index for $\{i\,\Lambda\,\mu\}$.
The normal-mode frequencies $\tilde\omega_j$ in each well are computed
by taking the square roots of the eigenvalues of the dynamical matrix
defined by
\begin{equation}
{\cal D}(n,S)_{\alpha \alpha' }=
\delta_{\alpha \alpha' } \, \omega_\alpha^2 + 
\omega_\alpha^{1/2}\,
\left[{\cal H}(n,S)_{\alpha\, \alpha' }
- \delta_{\alpha \alpha' } \omega_\alpha \right]
\, \omega_{\alpha'}^{1/2}.
\end{equation}
The $ - \delta_{\alpha \alpha' } \omega_\alpha$ term removes the restoring
term from $\hat{V}_{\rm vib}$ in $V^{\rm ad}_{n,S}$, and the
$\omega_\alpha^{1/2}$ are introduced by the standard change of variables to
correct for the different ``mass'' coefficients of different coordinates
$Q_{\alpha}$ in $\hat{T}_{\rm vib}$.
%
%
By retaining the harmonic expansion of the adiabatic potential around a
minimum 
the quantum ground state energy of this potential well is estimated at
$\sum_j\frac 12\hbar\tilde\omega_j$ above the classical minimum, due to
zero-point motion.
%
%
As illustrated in Fig.~\ref{zeroEnergetics:fig}, the difference between
this and the original zero-point energy $\sum_{\alpha} \frac 12
\hbar\omega_{\alpha}$ at the neutral-molecule harmonic minimum provides the
leading quantum correction
\begin{equation}
E_{\rm zero}(n,S) =
 \frac 12 \left[
\sum_j \hbar\tilde\omega_j (n,S) -\sum_{i\Lambda \mu}  \hbar \omega_{i\Lambda}
\right] ,
\label{Ezero}
\end{equation}
to the ``classical'' valley energy $E_{\rm adiab}$.
$E_{\rm zero}(n,S)$ in JT problems is systematically found to be negative
in sign, corresponding to softer vibrations, and a shallower valley bottom
at the new minima than for the $n=0$ non-JT molecule.
Table~\ref{energies:tab} reports the zero-point
correction computed as described above, and the estimate of the
ground-state energy obtained by adding the correction $E_{\rm zero}$ to
$E_{\rm adiab}$.

For C$_{60}^{n-}$, the zero-point corrections favor low-spin states for
both $n=2$ (already a $S=0$ ground state at the adiabatic level) and
$n=3$, where the zero-point correction only affects $S=1/2$, lowering 
it well below the uncorrected $S=3/2$.
Note however that in the $t^n\otimes H$ JT for the anions, the two ($n=1$ and
2) and three ($n=3$) ``soft'' modes along the trough are associated to
vanishing frequencies $\tilde\omega_j$. They formally contribute a 
vanishing zero-point energy to the first sum of
Eq.~(\ref{Ezero}). In our calculation this produces 
unreasonably large (negative) zero-point
corrections, larger even than the adiabatic energies.
These null terms are correct only in the limit of infinite coupling, where
the ``size'' of the flat trough is infinite, and indeed the free
pseudorotational motion carries no zero-point energy.
For C$_{60}^{n-}$, where coupling is finite, and not especially large, the
trough has a finite size of order $g^2$, which provides some amount of
quantum confinement, associated to a zero-point kinetic energy of order
$g^{-2}$, as was described in Ref.~\cite{AMT,ob96}.
These extra ``confinement'' corrections should be especially sizable when
the vibrational wavefunctions have the nontrivial nodal structure due to the
boundary conditions associated to an electronic Berry phase, i.e.\ for
$n=1$, $n=2$ $S=1$, and $n=3$ $S=1/2$ \cite{AMT}, but should not change the
conclusion that all C$_{60}^{n-}$ ions should favor low spin.

To get an estimate of the validity of the approximations employed here, 
we can compare with the ground state energy of C$_{60}^{-}$ obtained
by means of an essentially exact calculation (a Lanczos diagonalization 
on a truncated, but well converged, basis) of our Hamiltonian 
(\ref{modelhamiltonian}) with the same
parameters. The exact energy gain obtained is $-76$~meV, 
which falls in 
between the adiabatic ($-38$~meV) and 
zero-point corrected ($-152$~meV) values.
Not surprisingly, in this rather weakly coupled case, the zero-point
correction largely overshoots the correct value.
As charges $n>1$ are associated to larger distortions, thus effectively
to stronger couplings, we expect that 
the zero-point corrected energy should be of
better quality there.
However, the relative size of the zero-point and confinement corrections to
the adiabatic energy makes the whole strong-coupling expansion rather
questionable for all the anions. Thus the C$_{60}^{n-}$ results
as reported here serve mostly for comparison with those of the cations.
For cations we expect in fact the quality of our zero-point corrected
results to be substantially better.

In the C$_{60}^{n+}$ ions too, in fact, the zero-point energy 
corrections are rather large, but here the magnitude of the 
adiabatic energy gains is even larger.
The difference with the anions is due both to the larger 
couplings and to the structure of localized minima as opposed 
to a flat trough.
Comparison of Table~\ref{energies:tab} and Table~\ref{tab-symm} shows that
situations with a larger number of minima generally yield larger values of
the correction $|E_{\rm zero}|$.
Quantum kinetic energy in this respect behaves a bit like statistical
entropy, in that it favors numerous shallower minima against few deeper ones.
The zero-point correction peaks at $\approx -150$~meV for 
$n=5$ $S=\frac 12$, where the minima are shallow and the 
lowest vibrational frequency is as small as
$\tilde\omega_1\approx 10$~meV in this case.
On the contrary, few well spaced minima, as for $n=1$ $S=1/2$ and $n=2$
$S=0$ are associated to a smaller zero-point gain of the order $\approx
-100$~meV.
In the close competition between Coulomb physics (Hund's rules) and JT
physics (anti-Hund behavior), the zero-point correction is not
irrelevant. As
shown by the last column of Table~\ref{energies:tab}, it reduces
drastically the large adiabatic spin gap between the high-spin
ground state and the lowest spin excitations for $n=4$ and for $n=5$.
Remarkably, the zero-point correction favors the high-spin state instead 
in C$_{60}^{2+}$: the excitation energy to $S=0$ is
enhanced by the quantum correction from 13~meV to 33~meV.
Thus although the estimated spin gap is not especially large 
nor especially reliable, this  circumstance makes in our view 
the prediction of a $S=1$ magnetic C$_{60}^{2+}$ ion stronger
than for the other cations.

Could we go beyond the zero point correction, and get more
accurate results? Not easily at this stage.
The zero-point correction represents the $g^0$ term of a large-coupling
expansion, where the adiabatic energy $E_{\rm adiab}$ is the leading
($g^2$) term.
The next corrections to be considered, of order $g^{-2}$ and higher, are
associated with anharmonicity of the valleys and tunneling amongst them
(in the holes case), and also to mixing of the upper adiabatic potential
surfaces with the associated geometric-phase effects
\cite{AMT,hbyh,noberry,baer02}.
A full quantitative description of these effects would imply a much more
sophisticate treatment of the quantum problem than the simple semiclassic
expansion applied here, and that is beyond the scope of the present
paper.
To get an estimate of the importance of these higher-order corrections,
we can again compare, now for C$_{60}^{+}$  the adiabatic ($-69$~meV) 
and the zero-point corrected ($-120$~meV) ground-state energy gains 
to the exact on, obtained by a well converged Lanczos diagonalization, 
which is $-103$~meV. 
We see that although $E_{\rm zero}$ still overshoots the quantum 
corrections the adiabatic gain, the residual error of 17~meV is now
fairly small.
The accuracy expected for the zero-point corrected energy gains for all
other cations C$_{60}^{n+>1+}$ is even better. In fact, the distortions
there are larger, tunneling is suppressed, and the approximation of
isolated harmonic JT valleys should be much better than for $n=1$.
Thus, even if a check with the Lanczos method would be too cumbersome here because of
the excessive basis size required for $n>1$, our zero-point corrected
results of Table~\ref{energies:tab} should be regarded as practically
exact, for the assumed coupling parameters.

\rem{

Summary of the results up to now (all in meV):

\rem{
3 lowest vibron freq at minimum - all cases:
      1
n=1 S=- {31.2227, 31.2227, 31.2565}
      2
n=2 S=0 {30.2996, 30.2996, 30.4232}
n=2 S=1 {22.2793, 26.8695, 30.8925}
      1
n=3 S=- {18.7234, 26.4537, 30.3298}
      2
      3
n=3 S=- {22.2793, 26.8695, 30.8925}
      2
n=4 S=0 {13.9639, 13.964, 31.1871}
n=4 S=1 {12.0573, 29.4935, 30.2529}
n=4 S=2 {31.2227, 31.2227, 31.2566}
      1
n=5 S=- {9.69696, 19.6759, 29.4999}
      2
      3
n=5 S=- {29.0969, 29.4171, 30.5783}
      2
      5
n=5 S=- {32.4119, 32.4119, 32.4119}
      2
}

n=1 S=1/2\\
barrier height = 18.4687\\
Transverse-corrected barrier height = -27.147 (upside down potential)

n=2 S=0\\
barrier heigth = 32.6752\\
Transverse-corrected barrier height = 10.3872
}


\section{Discussion and Conclusions}
\label{discussion:sec}

The main output of this paper is a determination of the
ground state spin, energy, and distortion magnitude
of JT distorted C$_{60}^{n+}$ molecular ions. 

Calculations include both e-e and e-v interactions as derived
from earlier first-principles calculations, as well as
very important quantum vibrational effects, due to the small mass of
the carbon nuclei. The latter are taken into account 
approximately by including the changes of vibrational 
zero-point energy from the undistorted to the different distorted states.
These zero-point corrections are shown to be generally large.
In C$_{60}^{3-}$ they are even capable of turning a high-spin ground state
into a low-spin one.
%
%
As the coupling and thus the distortions are fairly large, quadratic and
higher-order (in ${\bf Q}$) e-v interactions and vibration anharmonicity
could also be relevant. The present calculation was however 
carried out strictly in the linear e-v coupling approximation.
%

The parameters used in this calculation, both for e-e and e-v interaction
could be somewhat underestimated by the local density approximation used in
their determination, as discussed in Ref.~\cite{Lueders02,Manini01}.
Consequently, both the Coulomb repulsion and the JT effective e-e
attraction calculated within the local density approximation might need
some correcting in their absolute values.
The balance between these two opposing interactions is delicate in
C$_{60}^{n-}$ ions (as demonstrated by the presence of both high-spin and
low-spin local ground states in different chemical environments
\cite{Brouet01,Arovas95,Kiefl,Zimmer95,Lukyanchuk95,Prassides99,Schilder94}),
but low-spin ground states are generally believed to prevail, 
in accord with the present calculation.
Contrary to that, in C$_{60}^{n+}$, Hund-rule magnetism 
and high-spin ground states are predicted to dominate.
Experimentally, so far we found no evidence concerning gas phase
C$_{60}^{n+}$ ions that we could usefully address. We do not know at this
moment whether the lifetime of positive ions against fragmentation would
permit experiments to be conducted, and if so whether the ground-state
spins and JT distortions could be determined and compared with our
predictions.
Singly charged C$_{60}^{+}$ ions have been created in solution
\cite{Reed00}, in molecular-beam photoemission \cite{Bruhwiler97,Canton02}, and
in storage rings \cite{Tomita01}: but they are irrelevant to our present
question.

More encouragingly, higher positive charge states have been 
pursued in solid-state acceptor compounds \cite{Datars95,Panich03}.  
The interplay of intra-molecular physics with electron hopping in a
hypothetical fullerene-cation based solid-state conducting compound is very
interesting. Our results suggest that magnetism should be important in
these compounds, at least so long as the hole bands remain relatively
narrow.
The HOMO bands of the proposed acceptor compounds (AsF$_6$)$_2$C$_{60}$ and
(SbF$_6$)$_2$C$_{60}$ can indeed be expected to be rather narrow.
In (AsF$_6$)$_2$C$_{60}$, based on a bct unit cell with lattice parameters
$a=b=12.8$~\AA, $c=12.4$~\AA, we can extract a fullerene-fullerene
inter-center distance 10.97~\AA, significantly larger
than 10.02~\AA\ of pure C$_{60}$.
On the other hand, the on-site Hubbard U estimated for holes is comparable
to or slightly larger than that for electrons.
It seems reasonable therefore to surmise that hole-doped
(AsF$_6$)$_2$C$_{60}$ could be, like the electron-doped K$_3$NH$_3$C$_{60}$
\cite{splitDMFT}, as
well as A$_4$C$_{60}$ \cite{Capone01,Capone02}, a Mott-JT insulator.
Unlike the electron-doped compounds, where the JT distorted C$_{60}^{n-}$
molecular ion was in a low-spin configuration ($S=1/2$ for $n=3$, $S=0$
for $n=2,4$), we expect in the divalent acceptor fullerene compound that the
positive C$_{60}^{2+}$ ions should be in a high-spin, $S=1$ JT-distorted
state. If so, they would constitute an exciting example of molecular Hund's
rule magnetism.
The magnetic coupling between neighboring C$_{60}^{2+}$ $S=1$ ions in these
compounds should most likely be weak and antiferromagnetic \cite{Reed00review}.
It might be interesting to pursue this and to seek a N\'eel state of some
kind at very low temperatures.
Pressure studies of the electron-doped compounds have revealed
insulator-metal transitions such as that of Rb$_4$C$_{60}$ \cite{Kerkoud},
and also insulator-superconductor transitions such as that of
K$_3$NH$_3$C$_{60}$ \cite{Margadonna01}.
Pressure studies of fullerene acceptors could be an interesting line to
pursue in the future.
More generally, the realization of a well characterized fullerene hole
conductor remains a worthy challenge for the future.

\section*{Acknowledgments}

We are indebted to M.\ Wierzbowska, G.\ Santoro, G.\ Onida, and L.\ Reatto
for useful discussions.
This work was supported by the European Union, contracts ERBFMRXCT\-970155
(TMR FULPROP), covering in particular the postdoctoral work of
M. L{\"u}ders; by HPRI-CT-1999-00048 (MINOS) and the CINECA Casalecchio
Supercomputing Center for computing time and for a fellowship;
and by MIUR CO\-FIN01, and FIRB RBAU017S8R of MIUR.

\appendix
\section{Appendix}

Tables~\ref{tab-modeM} and \ref{tab-modeP} report the individual amount of
JT distortion associated to each normal mode of C$_{60}$ at any equivalent
adiabatic minimum of C$_{60}^{n\pm}$, for all spin-$S$ sectors.
The distortions are given in the dimensionless units of the coordinate
operators $|\hat{Q}_{i\Lambda}|$.
The appropriate length scales $x_0(\omega_{i\Lambda}) =\sqrt{\hbar/
(\omega_{i\Lambda}\, m_{\rm C})}$ for the $G_g$ and $H_g$ modes of C$_{60}$
are: 76.3, 70.4, 60.3, 50.3, 46.1, 43.0, and 103.7, 80.9, 62.6, 59.8, 50.1,
47.0, 43.9, 42.1 pm, respectively.
For C$_{60}^{n-}$ the amount of ``radial'' distortion is independent of the
point chosen along the trough.
As expected, the modes characterized by the strongest coupling show the
largest distortion.
The mode associated to the largest distortion is therefore the lowest $H_g$
mode (see Table~\ref{e-v-params:table}) in the cations.
Note also that the $G_g$ modes do not contribute to the distortions of
$D_5$ symmetry, but contribute to all the lower-symmetry minima.

\begin{table}
\caption{\label{tab-modeM}
The dimensionless JT distortions $|\hat{Q}_{i\Lambda}|$ at the minima of
C$_{60}^{n-}$, for each $H_g$ mode and allowed spin $S$.
}
\begin{center}
\begin{tabular}{ccc}
\hline
\hline
$n$     &$S$ 		&       dimensionless distortions of $H_g$ modes\\
\hline
1       &       1/2     &       0.412 0.489 0.350 0.224 0.193 0.138 0.315 0.289 \\
2       &       0       &       0.813 0.965 0.691 0.442 0.381 0.273 0.622 0.571\\
2       &       1       &       0.412 0.489 0.350 0.224 0.193 0.138 0.315 0.289 \\
3       &       1/2     &       0.706 0.838 0.600 0.384 0.331 0.237 0.540 0.496  \\
3       &       3/2     &       0.000 0.000 0.000 0.000 0.000 0.000 0.000 0.000 \\
\hline
\end{tabular}
\end{center}
\end{table}

\begin{table*}
\caption{\label{tab-modeP}
The dimensionless JT distortions $|\hat{Q}_{i\Lambda}|$ at the minima of
C$_{60}^{n+}$, for each $G_g$ and $H_g$ mode and each allowed spin $S$.
}
\begin{center}
\begin{tabular}{cc|cll}
\hline
\hline
$n$     &$S$    & Symmetry &      distortions of $G_g$ modes  & distortions
of $H_g$ modes\\
\hline
1       &       0       & $D_{5d}$ &      0.000 0.000 0.000 0.000 0.000 0.000 
  &  1.36 0.473 .00459 0.35 0.023 0.205 0.379 0.333 \\
& & & \\
2       &       0       & $D_{5d}$ &      0.000 0.000 0.000 0.000 0.000 0.000
  &  2.69 0.935 .0095 0.692 .0455 0.405 0.749 0.657 \\
2       &       1       & $D_{2h}$ &      .0755 .0102 .0799 .0623 .0227 .0466 
  &  1.58 0.611 0.138 0.404 .0516 0.262 0.486 0.346   \\
& & & \\
3       &       1/2     & $C_{2h}$ &      .0548 .0074 .0580 .0452 .0165 .0339
  &  2.62 1.01 0.185 0.669 .0799 0.433 0.801 0.571    \\
3       &       3/2     & $D_{2h}$ &     .0755 .0102 .0799 .0623 .0227 .0466
  &  1.58 0.611 0.138 0.404 .0516 0.262 0.486 0.346   \\
& & & \\
4       &       0       & $D_{3d}$ &      .0828 .0112 .0877 .0683 .0249 .0512
  &  2.88 1.30 0.494 0.728 0.153 0.553 1.02 0.486       \\
4       &       1       & $C_{2h}$ &     0.074 0.010 .0782 0.061 .0223 .0457
  &  2.39 0.968 0.228 0.609 .0879 0.414 0.767 0.486   \\
4       &       2       & $D_{5d}$ &      0.000 0.000 0.000 0.000 0.000 0.000
  &  1.36 0.473 .0046 0.35 0.023 0.205 0.379 0.333    \\
& & & \\
5       &       1/2     & $C_{2h}$ &      0.101 .0135 0.106 .0827 .0302 .0619
  &  2.70 1.17 0.401 0.683 0.129 0.500 0.926 0.492        \\
5       &       3/2     & $C_{2h}$ &      .0384 .0053 .0411 0.032 .0117 0.024
  &  2.12 0.756 .0391 0.544 .0427 0.327 0.605 0.503  \\
5       &       5/2     & $I_h$ &      0.000 0.000 0.000 0.000 0.000 0.000
  &  0.00 0.000 0.000 0.000 0.000 0.000 0.000 0.000\\
\hline
\end{tabular}
\end{center}
\end{table*}



\begin{thebibliography}{10}
\bibitem{LeuvenError}
{ The treatment of zero-point corrections in Ref.\ \protect\cite{Leuven02} is
  incorrect. The present version gives the corrected formulas and values}.
\end{thebibliography}

\end{document}